\documentclass[
    ,final            
  ]
  {aipproc}
\usepackage{graphicx,amsmath,amssymb}
\layoutstyle{6x9}

\begin{document}

\title{Improved Higgs Naturalness with Supersymmetry}


\author{Ben Gripaios}{
  address={Rudolf Peierls Centre for Theoretical Physics, University of Oxford,
1 Keble Rd., Oxford OX1 3NP, UK} ,altaddress={Merton College, Oxford OX1 4JD, UK}
}

\author{Stephen M. West}{
  address={Rudolf Peierls Centre for Theoretical Physics, University of Oxford,
1 Keble Rd., Oxford OX1 3NP, UK}
}

\begin{abstract}
 We present a supersymmetric model of electroweak symmetry-breaking exhibiting improved naturalness, wherein the stop mass can be pushed beyond the reach of the Large Hadron Collider without unnatural fine tuning. This implies that supersymmetry may still solve the hierarchy problem, even if it eludes detection at the LHC.
\end{abstract}

\maketitle
\section{Constraints on the Scale of New Physics}
The standard model (SM) of particle physics provides an excellent fit to precision electroweak experiments, provided the mass $m_H$ of the as-yet unobserved Higgs boson is rather light \cite{unknown:2005em},
\begin{gather} \label{ewsm}
m_H \lesssim 285\  \text{GeV}\  (95\ \% \  \text{C.\ L.\ }).
\end{gather}
However, the SM suffers from the hierarchy problem: the electroweak scale is sensitive to higher energy scales occurring in nature, via a quadratically-divergent contribution to the Higgs mass parameter coming dominantly from virtual top quarks.  The divergence must be cut off by new physics at some energy scale $\Lambda_t$, which, in the absence of unnatural fine-tuning of parameters, should also be rather low.
This has been a major motivation for the ongoing construction of the Large Hadron Collider (LHC). A quantitative measure of the fine-tuning constraint may be given as \cite{Ellis:1986yg}
\begin{gather} \label{ftsm}
\Lambda_t \lesssim  400\ \text{GeV} \left( \frac{m_H}{115\ \text{GeV}} \right) \sqrt{D_H},
\end{gather}
where $D_H = \partial \log m_H^2 / \partial \log \Lambda_t^2$ and the amount of fine-tuning is roughly one part in $D_H$; in a natural theory, $D_H$ should be of order unity and the new physics will thus be accessible at the LHC.

The minimal supersymmetric extension of the standard model (MSSM), in which the quadratic divergence is elegantly cut off by the superpartners of the top quarks (the stops), is already unnatural in this sense: the stops must have masses of many hundreds of GeV in order to push the mass of the lightest Higgs boson above the empirical lower bound obtained from direct searches at LEP \cite{Inoue:1982pi} and the fine-tuning is at least a few {\it per cent}.

However, the electroweak (\ref{ewsm}) and fine-tuning (\ref{ftsm}) constraints are changed in a theory with an extended Higgs sector. 
Indeed, in such a model, the Higgs mass eigenstates that couple significantly to gauge bosons (and are thus constrained by electroweak precision tests) need not coincide with the eigenstates which are sensitive to $\Lambda_t$. The overlap between the different eigenstates (and {\it ergo} the constraints) is necessarily complete only in the theory with a single Higgs boson, {\it viz.} the SM. 
What is more, the individual constraints generalizing (\ref{ewsm}) and (\ref{ftsm}) are themselves changed in an extended Higgs sector and may also relax the constraint on the scale of new physics.

Recently, Barbieri and Hall \cite{Barbieri:2005kf} showed that even the simplest extension of the SM, the (non-supersymmetric) two Higgs doublet model, exhibits regions of parameter space with `improved naturalness', in which the combined electroweak and fine-tuning constraints are relaxed, allowing the scale of new physics to be as large as 2 TeV or more. This leads to the rather pessimistic conclusion that the new physics, whatever it may be, could lie beyond the reach of the LHC \cite{note2}. 

In the following, we exhibit a supersymmetric model of this type \cite{gripaios}, the so-called `fat Higgs model' \cite{Harnik:2003rs}. The existence of such a model implies that, even if no  direct evidence for supersymmetry is seen at the LHC, it cannot be ruled out as the solution to the hierarchy problem.
\section{A Supersymmetric Model with Improved Naturalness}
Models with supersymmetry necessarily have multiple Higgs doublets and are, as such, candidates for having improved naturalness. 
In supersymmetric models, the new physics (which cuts off the quadratic divergence in the Higgs mass parameters) is provided by superpartners.
The dominant divergence comes from top quarks, for which the contribution to the up-type Higgs mass parameter $m_2$ is given by
\begin{gather} \label{loop}
\Delta m_2^2 = - \frac{3 y_t^2}{4 \pi^2} m_{\tilde{t}}^{2} \log \frac{\Lambda}{m_{\tilde{t}}},
\end{gather}
where $y_t$ is the top quark Yukawa coupling, $m_{\tilde{t}}$ is the stop mass, and $\Lambda$ is the cut-off.
In a theory which remains perturbative up to, say, the GUT scale (such as the MSSM or the NMSSM), this relation implies a two-fold suppression of the natural stop mass.
Firstly, on the left-hand side, $m_2$ cannot be far above the weak scale, because it is related to the electroweak {\em vev} by a factor given roughly by the Higgs quartic couplings, which grow in the ultra-violet and so cannot be much larger than unity at the weak scale if they are to remain perturbative all the way up to the GUT scale.
Secondly, on the right-hand side, the logarithm (in which $\Lambda$ is the GUT scale) is large enough to cancel the loop factor. These two effects conspire in such a way that the superpartner scale is essentially the same as the weak scale, up to a factor of order unity.

The obvious way out is to discard the requirement that the theory remain perturbative up to the GUT scale. Then $m_2$ (and the Higgs masses) can be rather large, and the logarithm (in which $\Lambda$ is now the strong-coupling scale) is small.
What is more, successful unification of gauge couplings is not contingent upon quartic couplings remaining perturbative \cite{Nom,Harnik:2003rs}: It is possible to allow the theory to go through a calculable supersymmetric strong-coupling transition, whilst retaining gauge-coupling unification at higher energies (just as, for example, electroweak unification at around a hundred GeV
is not spoiled by the couplings of the low energy effective chiral Lagrangian for QCD becoming strongly-coupled at a GeV or so).

\begin{figure}
\includegraphics[width=7.5cm]{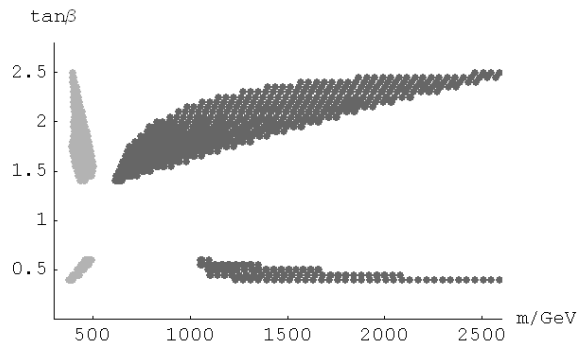}
\caption{\label{hb}Masses of the light (heavy) Higgs scalars {\it vs.} $\tan \beta$, where the light (dark) shaded regions have $m_{\tilde{t}}> 2$ TeV and satisfy precision electroweak tests at the 95\% C.L.. From \cite{gripaios}.}
\end{figure}

Below the strong coupling scale, the fat Higgs model \cite{Harnik:2003rs} contains the Higgs doublet chiral superfields $H_u$ and $H_d$ of the MSSM together with a singlet superfield $N$. The superpotential contains the terms 
\begin{gather} \nonumber
 \lambda N (H_d H_u - v_0^2) 
\end{gather}
 We expect $\lambda \simeq 4 \pi$ at the strong coupling scale $\Lambda$ on the basis of naive dimensional analysis \cite{Weinberg:1978kz}, while at a lower energy scale $\mu$, the coupling decreases according to the renormalization group.
Following \cite{Harnik:2003rs}, we take the quartic coupling $\lambda$ to be large compared to the gauge couplings $g$ and $g'$ and neglect the $D$-terms in what follows.
We have already given a qualitative explanation of how the fine-tuning constraint is alleviated in a strongly-coupled model. A rigorous discussion is given in \cite{gripaios}.
The fit to the electroweak data is described in \cite{gripaios}.
\begin{figure}
\includegraphics[width=7.5cm]{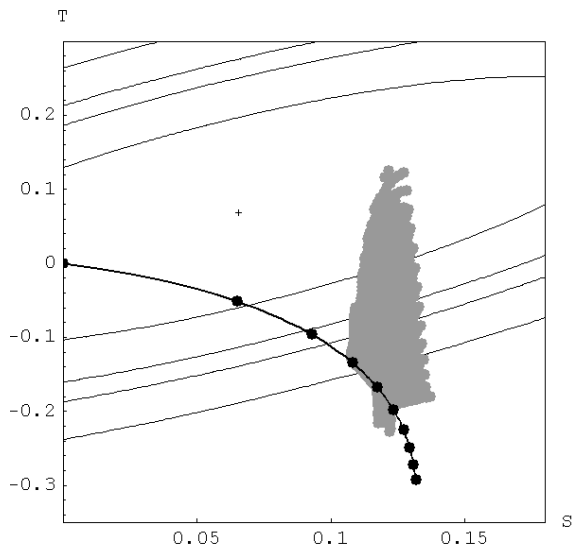}
\caption{\label{st}Constraints on the $S$ and $T$ parameters from electroweak data. The elliptical contours, centered on the best-fit values (+), enclose the 68, 90, 95 and 99 \% confidence regions. The grey area shows the model contribution for points in parameter space with $m_{\tilde{t}}> 2$ TeV and satisfying the $\Gamma (Z \rightarrow b\bar{b})$ and $b \rightarrow s\gamma$ constraints. For comparison, the thick line indicates the SM contribution with Higgs masses between 100 GeV (at the origin) and 1 TeV in increments of 100 GeV. From \cite{gripaios}.}
\end{figure}
Fig.~\ref{st} shows various confidence regions for $S$ and $T$ (relative to a SM Higgs reference mass of 100 GeV), together with the model contributions in regions of parameter space with natural stop mass above 2 TeV with the fine-tuning parameters $D_{h,H,Z} = \frac{\partial \log m^{2}_{h,H,Z}}{\partial \log m^{2}_{\tilde{t}}}\leq 4$, for the sake of argument), and satisfying the $\Gamma (Z \rightarrow b\bar{b})$ and $b \rightarrow s\gamma$ constraints. 
In Fig.~\ref{hb},
we plot the Higgs masses {\it versus} $\tan \beta$ for points with $m_{\tilde{t}}> 2$ TeV which are also compatible with electroweak constraints at the 95 $\%$ C.L. As is clear from Fig. \ref{st}, large values of $m_h$ are compatible with electroweak precision tests because there are large, positive contributions to $T$, coming from the other Higgs scalars. 
Can an extended Higgs sector of this type be discovered at the LHC? Both neutral Higgs scalars lie above the threshold for the `gold-plated' decay $H,h \rightarrow ZZ \rightarrow l \bar{l}l \bar{l}$ but the width is shared with the decay to $t \bar{t}$, which has a large background \cite{gun3}. Thus, it is not clear that either $h$ or $H$ will necessarily be discovered in this scenario. 



\bibliographystyle{aipproc}   

\end{document}